\documentclass[aps,showpacs,prb,twocolumn,floatfix,superscriptaddress]{revtex4}
\usepackage{amsmath}
\usepackage{amssymb}
\usepackage{graphicx}
\usepackage{dcolumn}
\usepackage{bm}

\begin{document}

\title{Pfaffian and fragmented states at $\nu={{5 \over 2}}$ in quantum Hall droplets}

\author{H. Saarikoski}
\affiliation{Kavli Institute of NanoScience, Delft University
of Technology, 2628 CJ Delft, the Netherlands}
\email[Electronic address:\;]{h.m.saarikoski@tudelft.nl}
\author{E. T\"ol\"o}
\affiliation{Helsinki Institute of Physics and Department of Applied
Physics, Helsinki University of Technology,
P.O. Box 4100, FIN-02015 HUT, Finland}
\author{A. Harju}
\affiliation{Helsinki Institute of Physics and Department of Applied
Physics, Helsinki University of Technology,
P.O. Box 4100, FIN-02015 HUT, Finland}
\author{E. R{\"a}s{\"a}nen}
\affiliation{Institut f{\"u}r Theoretische Physik,  
Freie Universit{\"a}t Berlin, Arnimallee 14, D-14195 Berlin, Germany}
\affiliation{European Theoretical Spectroscopy Facility (ETSF)}

\begin{abstract}
When a gas of electrons is confined to two dimensions,
application of a strong magnetic field may lead
to startling phenomena such as emergence of electron pairing.
According to a theory this manifests itself as appearance of
the fractional quantum Hall effect with a quantized conductivity
at an unusual half-integer $\nu={{5 \over 2}}$ Landau level filling.
Here we show that similar electron pairing may occur
in quantum dots where the gas of electrons is trapped by external
electric potentials into small quantum Hall droplets.
However, we also find theoretical and experimental evidence that,
depending on the shape of the external potential, the paired electron
state can break down, which leads to a fragmentation of charge and
spin densities into incompressible domains.
The fragmentation of the quantum Hall states could be an issue
in the proposed experiments that aim to probe for non-abelian
quasi-particle characteristics of the $\nu={{5 \over 2}}$ quantum Hall state.
\end{abstract}


\maketitle

\section{Introduction}

The discovery of the quantum Hall (QH) effect at Landau level filling 
factor $\nu={{5 \over 2}}$ in the two-dimensional electron gas (2DEG)
(Ref. \onlinecite{willett}) marked evidence that incompressible
states may form at unusual even-denominator filling fractions.
After years of subsequent theoretical and
experimental work\cite{nayakandsarma,pfaffi,greiter,morf}
it is well established that one of the most plausible theoretical
candidates for a QH state at $\nu=\frac{5}{2}$
is an exotic state of matter, a paired quantum Hall state.
Since electron-electron (e-e) interactions are repulsive
this pair formation is a collective phenomenon involving
residual interactions of composite particles 
that, in this state, are composites of an electron and two vortices.
The electron pairing would be analogous to the formation of Cooper pairs
in superconductors, although it would be purely a result of
e-e interactions without contribution from phonons or other fields.
In some theoretical models, the excitations of the paired electron state are
predicted to have non-abelian statistics that
could be employed in the field of topological quantum
computing.\cite{nayakandsarma}
Currently, the most pressing challenge is to experimentally find evidence
of the paired electron state and the particle statistics of its
excitations.\cite{radu,dolev,willett2}
The proposed tests\cite{sternbondersonfeldman}
for the non-abelian properties of these excitations
make use of confined geometries and multiple constrictions in the 2DEG
to generate interference among tunneling paths.
This leads to a natural question whether the paired electron state 
is stable when the 2DEG is confined into narrow trappings.

This work addresses the structure of the $\nu={{5 \over 2}}$ state
when electrons in the 2DEG
have been confined by external potentials
into small QH droplets.
They can be experimentally realized by placing semiconductor
quantum-dot (QD) devices into strong magnetic fields.\cite{reimann} 
We show here theoretical evidence that in QH droplets
the Pfaffian wave function,\cite{pfaffi}
which is commonly used to describe electron pairing,
may have high overlaps with the
exact many-body states at $\nu={{5 \over 2}}$.
In these calculations, we assume
that the half-filled Landau level is spin-polarized 
and use realistic e-e potentials that include
screening effects from the background charge of electrons in the 
the lowest Landau level and
a softening due to the finite thickness of the sample.
However, the half-filled second Landau level of
the Pfaffian state has a relatively high angular momentum,
which may lead to its instability in the QD confinement.
We show that in harmonic confining potentials
a compact filling of the half-filled Landau level is favored 
leading to the lowering of its angular momentum.
The paired electron state would then break down via fragmentation
of spin and charge densities into two incompressible domains,
spin-compensated $\nu=2$ at the edges and spin-polarized $\nu=3$
at the center (see Refs. \onlinecite{prlari} and \onlinecite{spindroplet}).
This phenomenon is analogous to the proposed
formation of similar structures in the 2DEG where translational symmetry
has been broken by long-range disorder.\cite{chklovskii}
We present the fragmented states in QDs as
alternatives to the Pfaffian state and show signatures
of them in electron transport experiments. 
Based on these results, we conjecture that the stability
of the paired electron state depends crucially on the shape
of the potential landscape where the electrons move in the 2DEG. This
may explain, e.g., the observed fragility of the $\nu={5 \over 2}$
QH state in narrow quantum point contacts.\cite{miller}

The paper is organized as follows. We introduce our theoretical model of
QDs in Sec.~\ref{sec:model}
and the computational methods used to solve the many-body
problem in Sec.~\ref{sec:methods}. The exact diagonalization
method is used in Sec.~\ref{sec:pfaffi} to calculate
the overlaps of the Pfaffian wave function with the exact
many-body state. In Sec.~\ref{sec:spin-droplet}, we analyze
the electronic structure of fragmented QH states and show
that the second-lowest Landau level is spin-polarized
due to the lifting of degeneracy of single-particle states near
the Fermi level. In Sec.~\ref{sec:experiments}, we present experimental
evidence for fragmentation of QH states
in the $2\le\nu\le{5\over 2}$ filling-factor regime.
Section \ref{sec:conclusions} concludes our work with discussion
of the relevance of our findings with the observed fragility of the
$\nu={5\over 2}$ QH state in disordered or confined 2DEG.

\section{Model}

\label{sec:model}
QDs formed 
in the ${\rm GaAs}/{\rm Al}_{x}{\rm Ga}_{1-x}{\rm As}$
heterostructure are modeled for both lateral and vertical QD
devices as droplets of electrons in a strictly two-dimensional (2D)
plane confined by a parabolic external potential.\cite{reimann}
We use an effective-mass Hamiltonian 
\begin{equation}
H=\sum^N_{i=1}\left[
 \frac{({\bf p}_i+e {\bf A} )^2}{2 m^*}
+V_{{\rm c}}(r_i)\right] + \frac{e^2}{4\pi \epsilon} \sum_{i<j}
\frac{1}{r_{ij}},
\label{hamiltonian}
\end{equation}
where $N$ is the number of electrons,
$V_{{\rm c}}(r)=m^*\omega_0^2 r^2/2$ is the
external parabolic confinement, $m^*=0.067m_e$ is the effective mass and
$\epsilon=12.7\epsilon_0$ is the dielectric constant
of the GaAs semiconductor medium, and ${\bf A}$ is the vector potential
of the homogeneous  magnetic field ${\bf B}$ perpendicular to the QD plane. 
The confinement strength $\omega_0$ in the calculations is $2\,{\rm meV}$,
unless otherwise stated.

If the e-e interactions are excluded, the single-particle solutions of the
Hamiltonian (\ref{hamiltonian}) are Fock-Darwin states.\cite{fockdarwin}
In the limit of a very high magnetic field, the Landau level structure
approaches that of the 2DEG. However, in finite magnetic fields
the external potential
alters the electronic structure and different Landau levels overlap.
Therefore, the concept of Landau level filling needs
to be generalized to finite-size systems. Kinaret {\it et
al.} defined the average filling factor as
$\nu_{\rm avg}=N^2(N+L)/2$, where $L$ is the total
angular momentum.\cite{kinaret}
Another possibility is to focus on the lowest Landau level (LLL)
and define filling factor of a state as $\nu_{\rm LLL}=2N/N_{\rm LLL}$.
These definitions differ in the high filling-factor regime, but
this is not critical to the interpretation of results that are
based on the structural properties of the many-body states.

\section{Computational many-body methods}

\label{sec:methods}
The ground state corresponding to interacting electrons in QH
droplets is solved numerically using the exact diagonalization (ED),
density-functional theory (DFT), and the variational quantum Monte Carlo
method (QMC). Since the paired electron state in the 2DEG is a strongly
correlated many-body state, the ED method is used to analyze
its stability in the QD confining potential.
The DFT and QMC methods
are used to analyze the fragmented QH states.
The regime where this fragmentation gives characteristic signals
in the experiments is beyond the reach of the ED method.
However, we find that both the DFT and QMC methods
provide accurate results in this regime (see Appendix).

\subsection{Exact diagonalization}

In the ED method, we assume that the electrons occupy states on one Landau
level only.  If we now take a fixed number of states from this Landau
level, our computational task is first to construct the many-body
basis.  Then the Hamiltonian matrix corresponding to Hamiltonian of
Eq.~(\ref{hamiltonian}) is constructed in this basis. Finally, the
lowest eigenstate and eigenvalue are found by matrix diagonalization.
More details can be found, e.g., in Ref.~\onlinecite{sigga}. In
addition to the standard Coulomb interaction, we use in the ED two
modifications of it. To model the finite thickness of the sample,
we use a softened potential~\cite{sarma} defined as
\begin{equation}
V_T(r)=\frac{e^2}{4 \pi \epsilon \sqrt{r^2+d_T^2}} \ ,
\label{V_T}
\end{equation}
where $d_T$ is the sample thickness. Electrons in second or higher Landau
levels move on top of background charge of lower Landau levels, which
effectively screens the Coulomb interaction. This is modeled
with a screened potential that is of the Gaussian form
\begin{equation}
V_S(r)=\frac{e^2\exp({-r^2/d_S^2})}{4 \pi \epsilon r} \ ,
\label{V_S}
\end{equation}
where $d_S$ is the screening length.
The unit of length in our ED results is given by
$l=\sqrt{\hbar/m^*\omega}$, where
$\omega=\sqrt{\omega_0^2+(\omega_c/2)^2}$ and
$\omega_c=eB/m^*$ is the cyclotron frequency of electron in magnetic field $B$.

\subsection{Density-functional theory}

Our DFT approach is based on spin-DFT, a variant of the 
conventional DFT generalized to deal with
non-zero spin polarization. On top of standard spin-DFT, 
we include the bare external vector potential ${\mathbf A}$ 
[see Eq.~(\ref{hamiltonian})] in the Kohn-Sham equation. 
In contrast with current-spin-DFT, however,
we neglect the exchange-correlation vector potential ${\mathbf A}_{xc}$.
In the magnetic-field range considered here, this has been shown 
to be a very reasonable approximation.~\cite{helbig}
As another valid approximation, 
we neglect the dependence of the exchange and correlation 
on the vorticity.\cite{testing} 
The exchange and correlation energies and potentials are
calculated using the 2D local spin-density approximation, 
for which we use the QMC parametrization of the correlation energy
by Attaccalite {\em et al.}\cite{attaccalite}

The DFT approach is implemented on a 2D real-space grid and 
employs a multigrid method for solving of the Kohn-Sham
equations.\cite{implementation}
Our symmetry-unrestricted DFT approach has been shown to lead to solutions
with broken rotational symmetry
that has been linked to mixing of the
different eigenstates of angular momentum.\cite{hirose,broken}
In a fixed symmetric external potential, this type
of spontaneous symmetry breaking is expected to be unphysical.
In Sec.~\ref{sec:experiments}, we compare the validity of this
assumption directly to experimental data.

\subsection{Quantum Monte Carlo}

Since the fragmentation of many-body state in the vicinity of $\nu={5 \over 2}$
is a delicate many-body problem, we
employ the QMC method to analyze the reliability of our
DFT approach. The wave function in the QMC is chosen to be
\begin{equation}
\Psi = D_{\uparrow} D_{\downarrow} \prod_{i<j}^N J(r_{ij}) \ ,
\label{wf}
\end{equation}
where the two first factors are Slater determinants for the two spin
types, and $J$ is a Jastrow two-body correlation factor.
The Slater determinants are constructed from the Fock-Darwin states.
For the two-body Jastrow factor, we use a form
\begin{equation}
J(r)=\exp\left({\frac{C r}{a+b r}}\right) \ ,
\label{Jsimple}
\end{equation}
where $a$ is fixed by the cusp condition to be three for a pair of equal
spins and one for opposite ones, and $b$ is an additional parameter 
different for both spin-pair possibilities. 
The ground state of the QD in the spin-droplet regime
is calculated assuming that the LLL and the
second-lowest Landau level (SLL) are compact.
This means that the Slater determinants are built from
single-particle states having angular momenta 
$l=0,\ldots,N_{{\rm LLL},s}-1$ for
spin $s=\uparrow,\downarrow$
in the LLL and $l=-1,\ldots,N_{{\rm SLL},s}-2$ for the spin $s$
in the SLL. Energy for each combination of non-negative total spin 
$S$ and total angular momentum $L$ is then
calculated. The QMC method deals with the correlation effects in the
many-particle system more accurately than the DFT approach. 
However, the computational cost of the QMC is significantly
larger than that of the DFT.
A detailed description of the QMC method
is given in Ref.~\onlinecite{AriLowTempPhys}.

\section{Pfaffian state in quantum dots}
\label{sec:pfaffi}

The structure of the QH states in the 2DEG at half-integer filling
factor has been a topic of intense research efforts.\cite{nayakandsarma}
Currently, it is regarded plausible that the experimentally observed
$\nu=\frac{5}{2}$ state consists essentially of a full spin-compensated
LLL and a half-filled spin-polarized SLL,\cite{morf} in which weak $p$-wave
electron pairing takes place.  Formally, the SLL is described by a
Moore-Read, or Pfaffian, wave function lifted to the
SLL.\cite{pfaffi,greiter}
There exists some theoretical evidence that the
excitations of this QH state obey non-abelian
statistics.\cite{nayakandsarma,pfaffi,nayak}
ED calculations
have become standard tests of trial wave functions of QH
states, and they have shown high overlaps with the Pfaffian wave
function in the 2DEG.\cite{rezayi}
However, there are other candidates for the $\nu={{5 \over 2}}$ state, some of
which possess only abelian quasiparticle excitations.\cite{nayak,overbosch}

The structure of the $\nu={1\over 2}$ state in QDs
was analyzed with the ED method in
Ref.~\onlinecite{prlari}.
Here we provide results for half-filled higher Landau levels 
with more realistic inter-electron potentials defined in
Sec.~\ref{sec:methods}.
Following the theory of the $\nu={5 \over 2}$ QH
state in the 2DEG, we assume that the half-filled Landau level is spin-polarized.
The Pfaffian wave function,\cite{pfaffi} which describes
paired fermion
states of the half-filled Landau level, is defined for LLL as
\begin{equation}
 \Psi_{\rm PF} = \mathrm{Pf} \left( \frac{1}{z_i-z_j} \right) \prod_{i<j} (z_i-z_j)^2
\exp\left(-\frac 12 \sum_i r_i^2\right).
\label{eq:pfaffi}
\end{equation}
In higher Landau levels the Pfaffian state is obtained
by applying the Landau level raising operator to each electron.
The angular momentum of the Pfaffian state is
$L'=N'(N'-1)-(n_{\rm LL}+{1\over 2})N'$,
where $N'$ is the number of electrons
in the half-filled Landau level and
$n_{\rm LL}=\{0,1,\ldots\}$ is the Landau level index.

We present the overlaps of the Pfaffian wave function with the ED
eigenstate for electrons frozen to lowest (LLL), second (SLL), or
third (TLL) Landau level, which correspond to filling fractions
of $\nu={1\over 2}$, $\nu={{5 \over 2}}$, and $\nu={9 \over 2}$,
respectively.
Electrons in the half-filled second and third Landau levels
move on top of the uniform background electron density of the
spin-compensated lower Landau levels. This background charge
effectively screens the Coulomb interaction. In QDs, the
e-e interactions are further screened due to metallic leads.

\begin{figure}
\includegraphics[width=0.99\columnwidth]{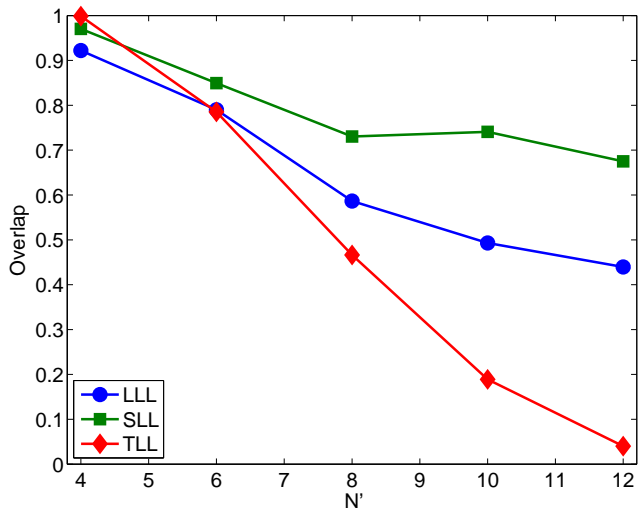}
\caption{(Color online) The overlaps of the Pfaffian wave function
with the corresponding exact state at the lowest (LLL), second
(SLL), and third (TLL) Landau level in the case of a Coulombic
electron-electron interaction. $N'$ denotes the number of
electrons in the half-filled Landau level.}
\label{fig:overlaps0}
\end{figure}

Figure~\ref{fig:overlaps0} shows the overlaps of the Pfaffian wave
function and the ED eigenstate of Coulomb interaction for particle
numbers $4\le N'\le 12$. For large particle numbers,
the overlaps in the second Landau level are highest. This shows that
$\nu={5 \over 2}$ has the highest probability to be described by the Pfaffian.

\begin{figure*}
\includegraphics[width=0.75\columnwidth]{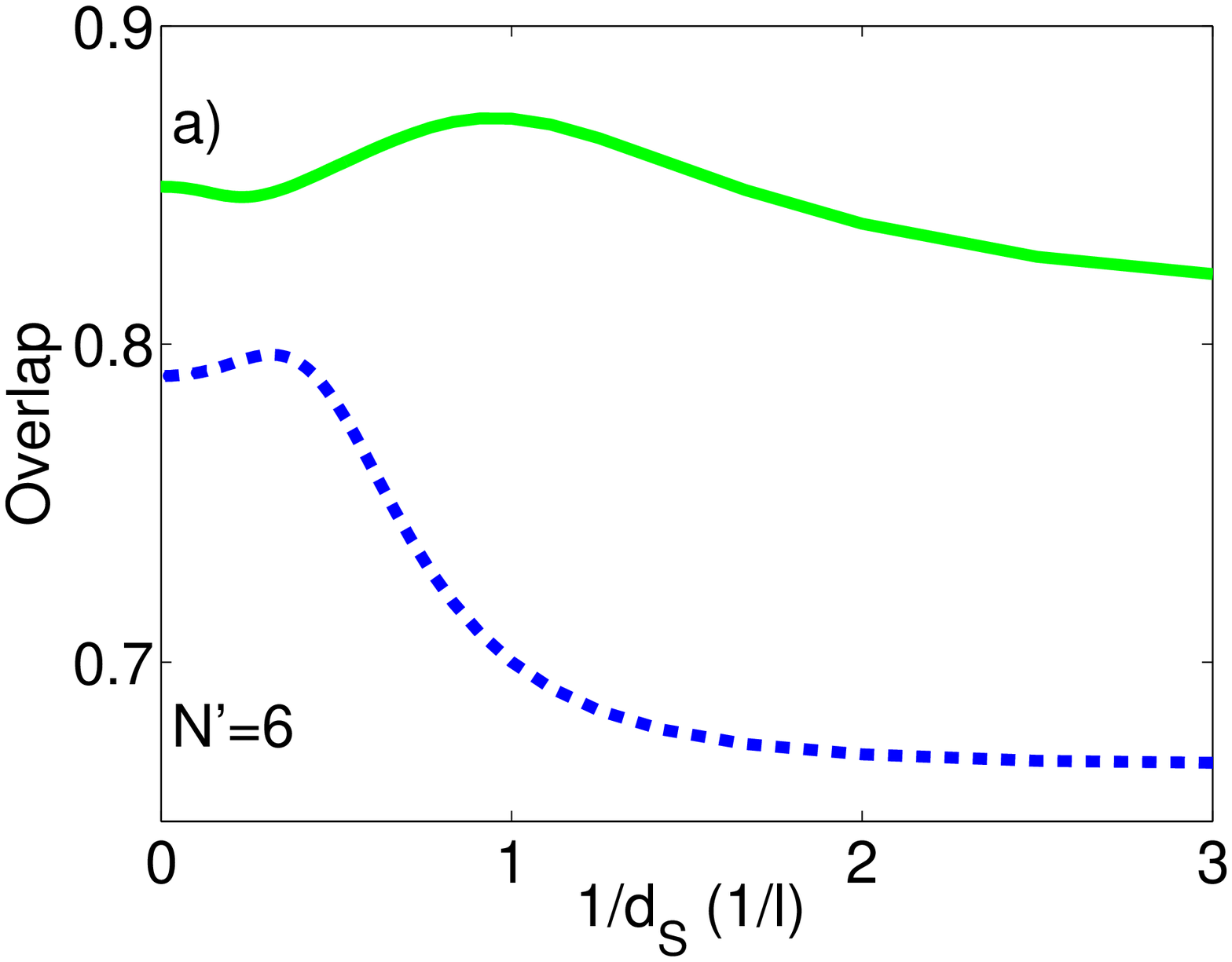}
\includegraphics[width=0.75\columnwidth]{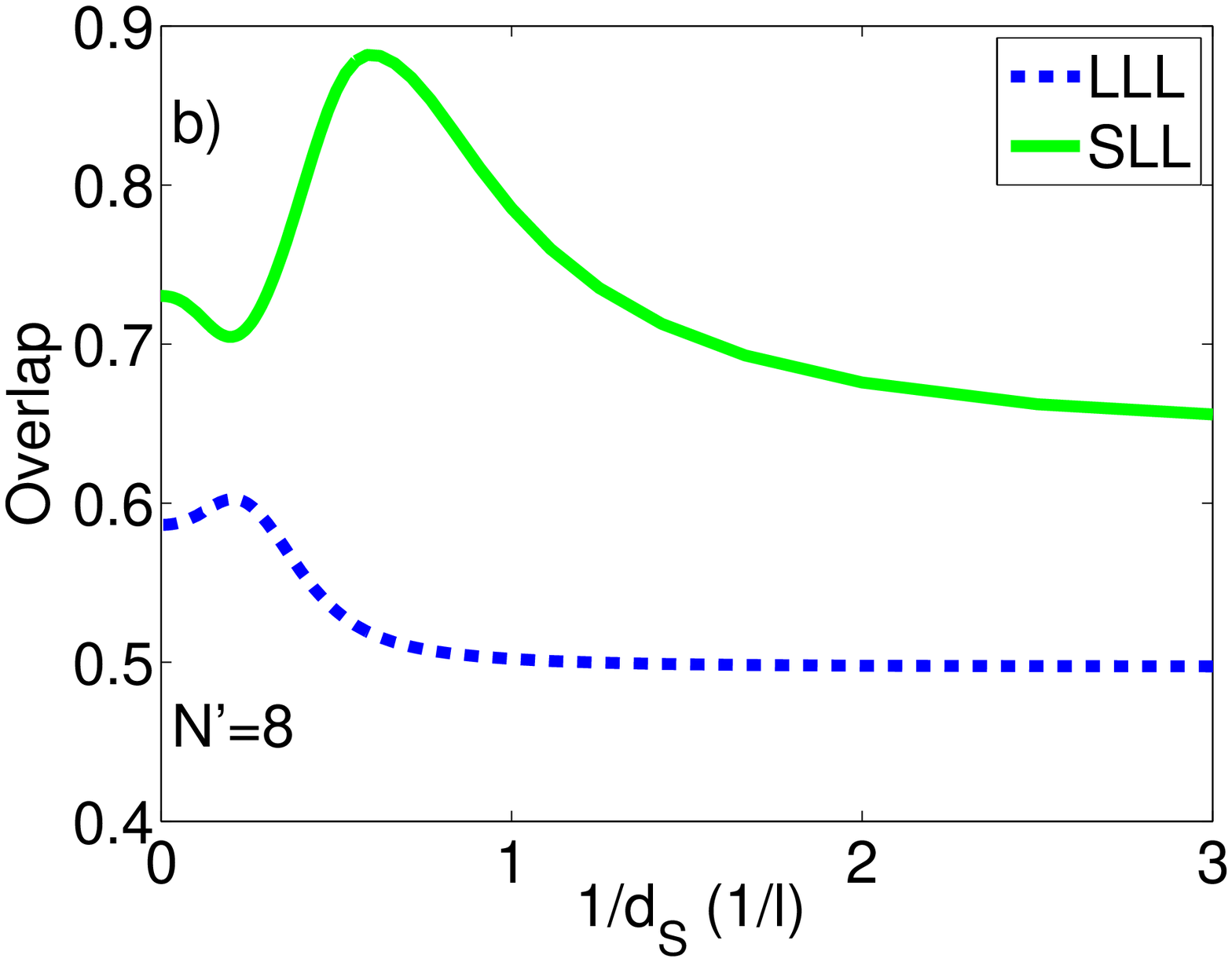}
\includegraphics[width=0.75\columnwidth]{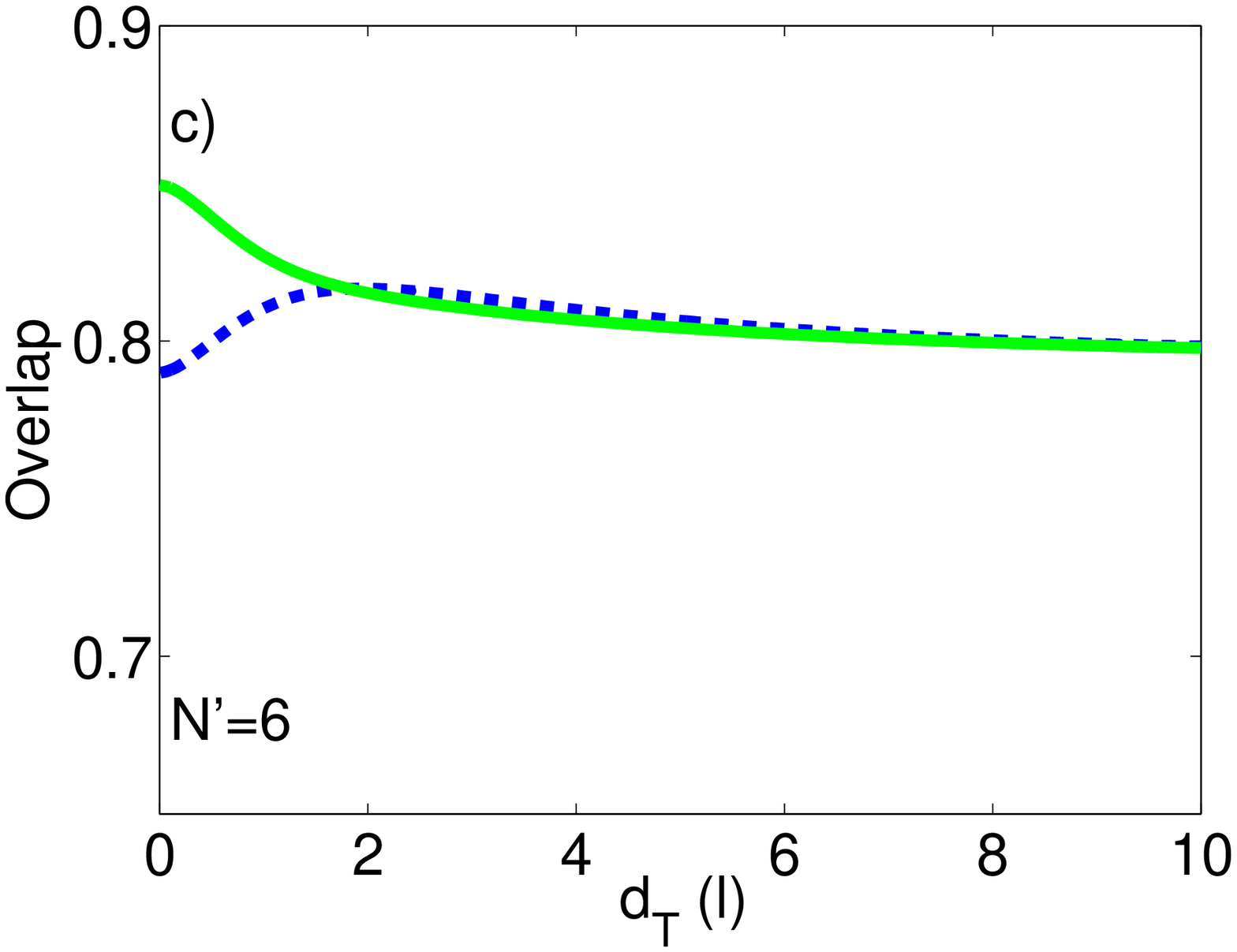}
\includegraphics[width=0.75\columnwidth]{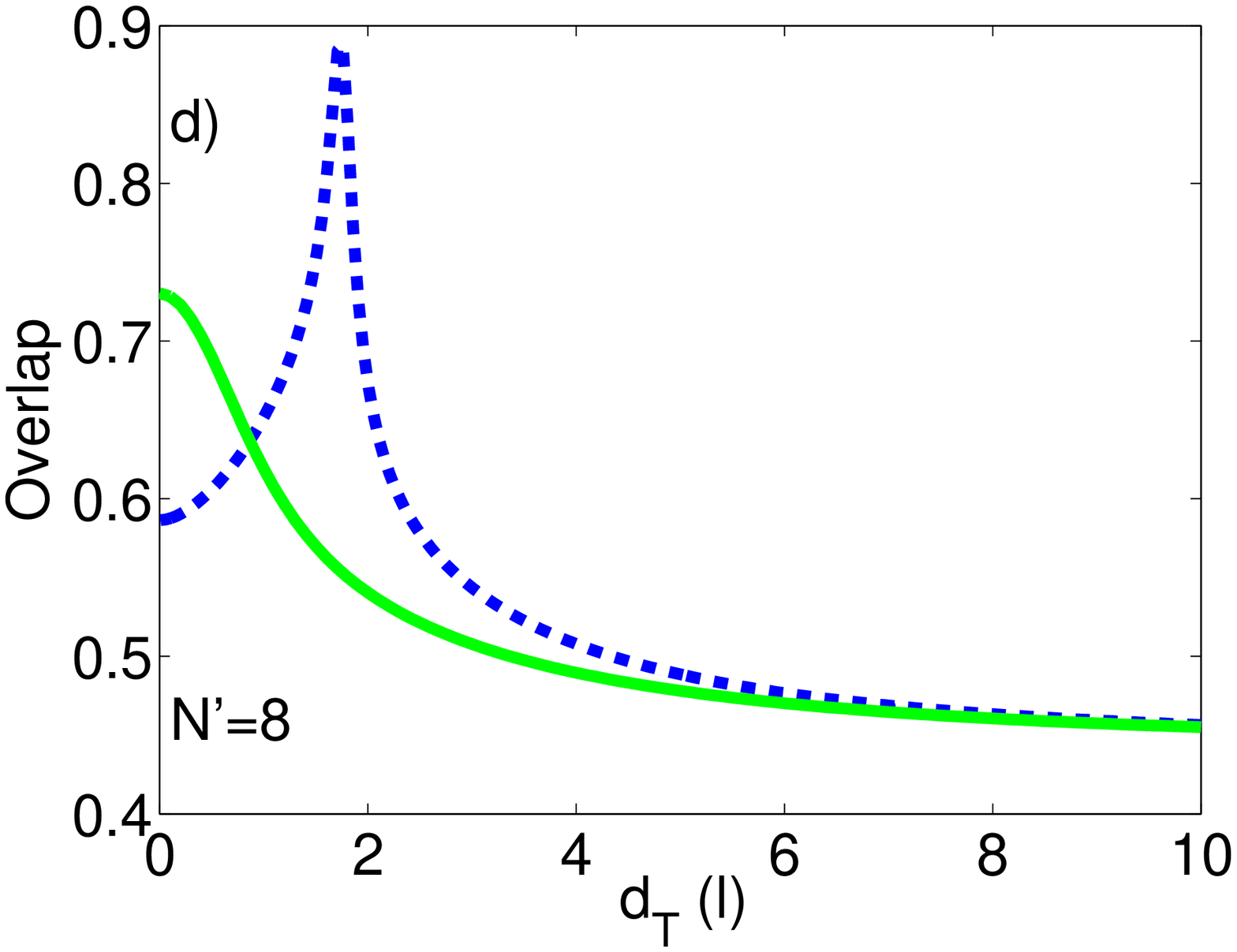}
\caption{Overlap of the Pfaffian wave function with the
corresponding exact state for $N'=6$ and $N'=8$ electrons
in the half-filled Landau level, respectively, using (a-b) screened
electron-electron potential with screening length $d_S$
and (c-d) softened potential due to finite sample thickness $d_T$
for electrons at the lowest Landau level 
(LLL) corresponding to $\nu={1\over 2}$,
and the second lowest Landau level (SLL) corresponding to $\nu={5\over 2}$.}
\label{fig:overlaps68}
\end{figure*}

Next, we study how the screening of the e-e interaction
and finite thickness of the sample
change the overlaps of the ED eigenstate with the Pfaffian.
For six electrons on
LLL, the overlaps are slightly improved when the screening and finite
sample thickness are taken into account in the interaction
(see Fig.~\ref{fig:overlaps68}). On SLL,
screening slightly improves the overlap, but a finite thickness lowers it.  The
same trends can be seen in Fig.~\ref{fig:overlaps68} for eight
electrons, but now the effects are clearly stronger, and there is a
large increase in the overlaps.
On the LLL, a finite sample
thickness is needed to achieve the best overlap. On the SLL,
the screening increases the overlap, which can be contrasted with
the spherical geometry where the SLL overlap
is maximized at a finite thickness of the sample.\cite{Peterson}

The highest overlaps are of the order of 0.8--0.9 at $\nu={5\over 2}$,
which means that the structure of the many-body state is close to the
Pfaffian. The exact state at the given angular momentum would therefore
show electron pairing to a significant degree.
We note that the Pfaffian wave function in Eq. (\ref{eq:pfaffi})
has no adjustable parameters. It is possible to 
modify the Pfaffian wave function by introducing a pairing function
that differs from $g=1/(z_i-z_j)$ of the Moore-Read form.\cite{moller}
This won't change the angular momentum of the state but has been
found to increase overlaps significantly in the 2DEG.

 In addition to the overlaps, it is crucial to verify
that the state at the angular momentum of the Pfaffian state is
energetically favorable.
In fact, the LLL $\nu=\frac{1}{2}$ state with
$N'=8$ corresponding to Fig. \ref{fig:overlaps68}(d) is a possible ground
state at small and large values of the thickness $d_T$, but not at values
of $d_T$ where the overlap is peaked.\cite{toloharju}
A further obstacle for the Pfaffian state in
finite-size QH droplets is that the SLL may not attain the high angular
momentum and complete spin polarization of the Pfaffian.
In QH droplets, the degeneracy of Landau levels is lifted
when electrons move in external confining potentials
[Fig. \ref{supplyfig4}(b)], and a
compact distribution of electrons on the Landau levels
could be energetically more favorable.
In the next section, we show that this would lead to nonexistence
of the paired electron state and introduce fragmented QH states
in quantum Hall droplets as alternatives.

\section{Fragmented quantum Hall states}
\label{sec:spin-droplet}

In quantum Hall droplets, single-particle
states within each Landau level are not degenerate
due to the confining potential.
The average distance of an electron
from the center of the droplet, and therefore also the potential energy,
increase with angular momentum. This suggests that
a compact occupation structure may be energetically favorable.
The compact occupation of Landau levels leads to fragmentation
of charge and spin densities into incompressible integer filling
factor domains. We call these states fragmented quantum Hall states
that are alternatives to the paired electron state at half-integer
Landau level fillings.

\begin{figure}
\includegraphics[width=0.99\columnwidth]{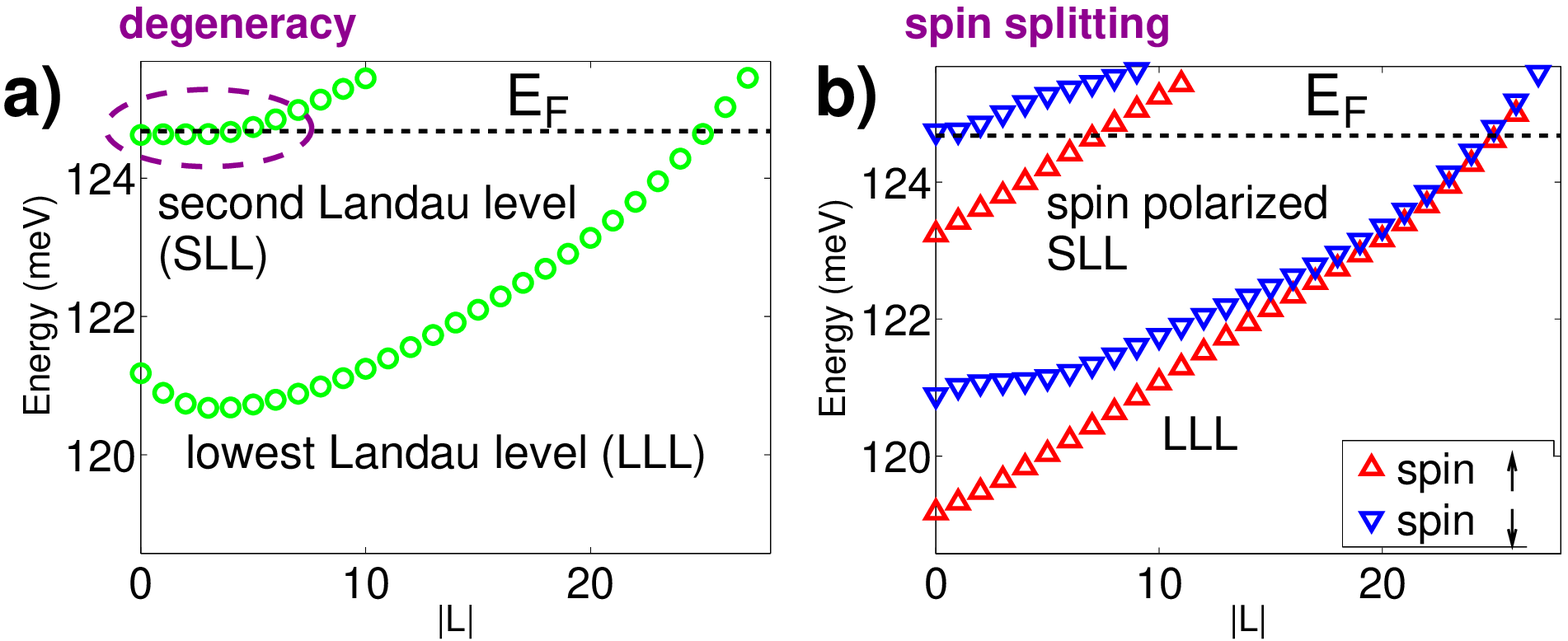}
\caption{(Color online)
(a) Kohn-Sham energy spectrum of a 60-electron quantum dot
as a function of single-particle angular momentum $L$ calculated
from the density functional theory with spin-compensated orbitals.
The density of states of the second-lowest Landau level (SLL)
is high near the Fermi energy $E_{\rm F}$. The magnetic field is 2.125 T
which corresponds to a filling factor of $\nu={{5 \over 2}}$.
(b) The corresponding energy spectrum from spin-density functional theory
shows lifting of the degeneracy near the Fermi level
via complete spin polarization of the SLL.
The lowest Landau level (LLL) remains spin compensated.
Spin $\uparrow$ ($\downarrow$)
corresponds to spin orientation parallel (antiparallel) to the magnetic field.
The spin-splitting due to many-body effects is about 1.5 meV at $L=0$.
In comparison, the Zeeman splitting is about 0.05 meV.
}
\label{supplyfig4}
\end{figure}

\begin{figure}
\includegraphics[width=0.85\columnwidth]{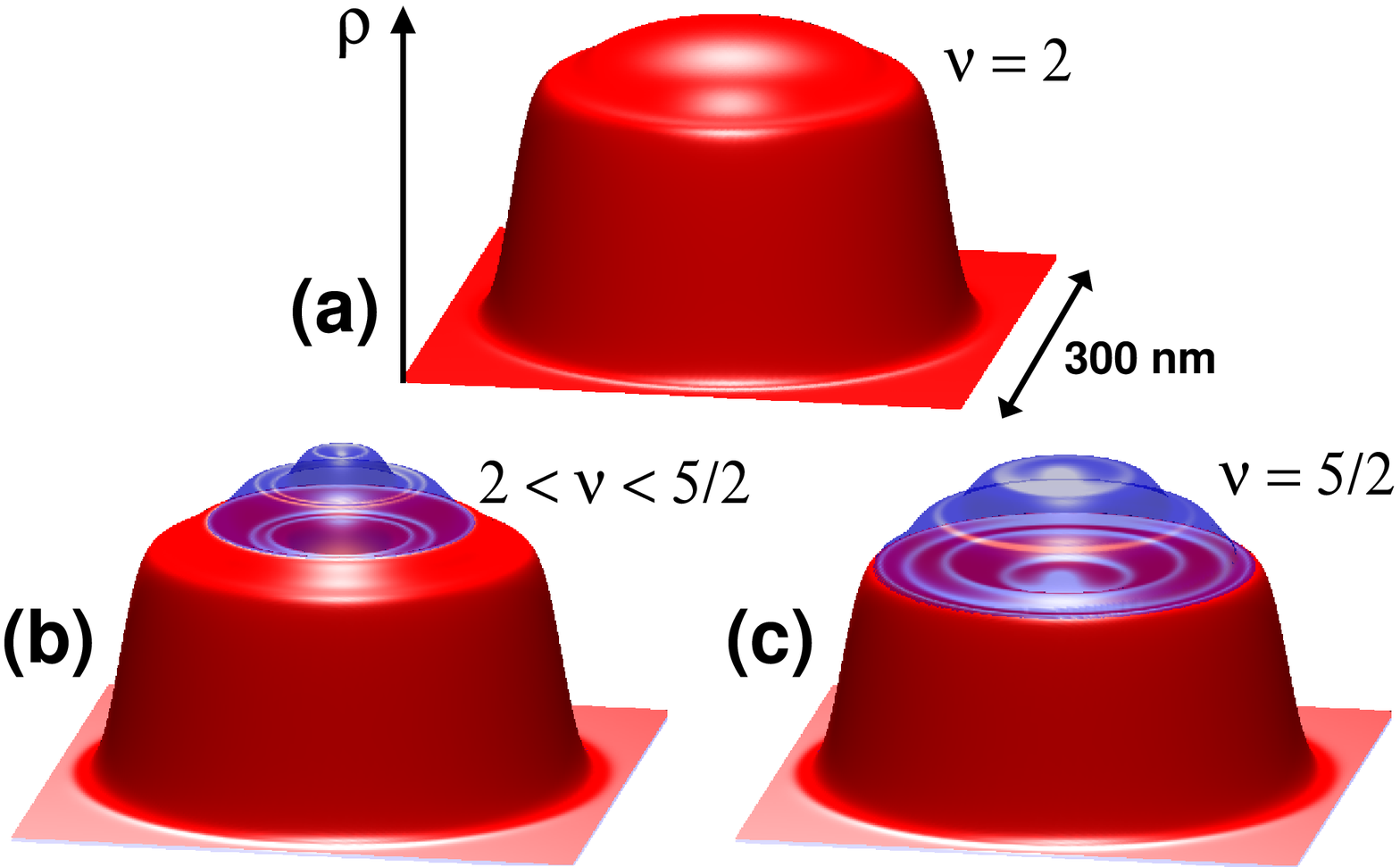}
\caption{
(Color online) 
Total electron density $\rho_\uparrow+\rho_\downarrow$ (full region) and
the net spin density 
$\rho_\uparrow-\rho_\downarrow$ (transparent blue region)
of quantum Hall states in a quantum dot
at (a) $\nu=2$, at (b) an intermediate state
between $\nu=2$ and $\nu={5 \over 2}$, and at (c) $\nu={5 \over 2}$. 
The latter two show fragmented charge and spin densities
with spin-compensated $\nu=2$ region at the edges and spin-polarized
$\nu=3$ at the center.
The densities were calculated with the spin-density-functional theory
for a 60-electron quantum dot.
The net spin-up density is due to
spin polarization of the second-lowest Landau level.
}
\label{tippafigure}
\end{figure}

We analyze the structure of fragmented QH states near $\nu={5 \over 2}$
in a harmonic confining potential of a semiconductor
quantum dot with the QMC and the DFT methods.
The Kohn-Sham single-particle energy spectrum of the 
Landau levels calculated with the spin-compensated DFT
and the spin-DFT are shown in Fig.~\ref{supplyfig4}(a) and (b), respectively.
The spin-DFT and the QMC show that the degeneracy of the 
single-particle states close the Fermi energy is lifted
via a complete polarization of the second-lowest Landau level.
Therefore, a compact occupation of the single-particle states of the 
spin-compensated LLL and spin-polarized SLL leads to a
fragmented state with a $\nu=2$ region (double-occupied LLL) at the
edges of the droplet and $\nu=3$ (spin-polarized SLL) at the center
[Fig.~\ref{tippafigure}].

The spin-splitting of the SLL in the spin-DFT calculations is
analogous to the Stoner criterion,
which states that in the presence of correlations between
electrons of the same spin and high density of states near
the Fermi level, the system prefers ferromagnetic alignment
that reduces the degeneracy.\cite{stoner}
We call the incompressible, spin-polarized droplet of SLL electrons
a {\em spin droplet}.\cite{spindroplet}
The size of the spin-droplet becomes significant when the number of electrons
in the dot $N\gtrsim 35$.
The non-uniform filling-factor structure of the spin-droplet states
is reminiscent of the incompressible QH domains
that form in the 2DEG with long-range disorder.~\cite{chklovskii}
The compact occupation of the SLL leads to a lower angular momentum
than what is needed for a paired
electron state as described by the Pfaffian wave function (\ref{eq:pfaffi}).
For example, the size of the spin-droplet in the QMC method 
is $N'=8$ electrons at $N=48$, and the angular momentum of the 
SLL is $L'=20$, which can be contrasted to $L'=44$ for the Pfaffian wave
function with the same number of electrons.

The SLL remains polarized and compact between ${{5 \over 2}}\ge \nu\ge 2$.
Hence, we call this filling-factor range the spin-droplet regime.
The size of the spin-droplet gradually shrinks with the increasing magnetic
field as the electrons are passed from the SLL to the LLL.
The contributions of the LLL and SLL occupancies to the electron
and spin densities are shown in Fig.~\ref{tippafigure} for the
case of 60-electron QD.
Qualitatively similar results were obtained for confinement
strengths 1 to 4 meV and electron numbers $N$ between 35 and
120, which confirms the generality of the results.
The calculations show that the energy benefit from
the polarization of the SLL is large (see Fig. \ref{supplyfig4} and Appendix),
which would make spin-droplets robust in the presence of 
impurities in samples.

We note that the stability of the fragmented QH states
in large QDs $(N>30)$ can be contrasted to the
instability of the maximum-density-droplet (MDD)
state in the same regime. The MDD state 
is the totally polarized state corresponding to the $\nu=1$
QH state in 2DEG, and it has been found to be 
instable in large QDs with
$N>30$~(Refs. \onlinecite{mddstability} and \onlinecite{oosterkamp}).

\section{Signatures of fragmentation in electron and spin transport}
\label{sec:experiments}

The emergence of finite size counterparts of integer
and fractional QH states in QDs gives characteristic
signatures in the chemical potentials. Several experimental methods
have been developed to measure the chemical potential in a QD via
addition of electrons one-by-one into the system. 
These experimental methods include Coulomb blockade,\cite{mceuen}
capacitance,\cite{ashoori} and charge detection techniques.\cite{field}
In this work, we use data from electron transport measurements of QDs
in the Coulomb and spin blockade regime.\cite{oosterkamp,ciorga00}
The spacings of the spin and Coulomb blockade peaks correspond
to the energy needed to add the $N$th electron in
the system of $N-1$ electrons, i.e., the chemical potential defined 
as $\mu(N,B)=E_{\rm tot}(N,B)-E_{\rm tot}(N-1,B)$.

We calculate the signatures in the chemical potentials associated
with the formation of fragmented QH states
and compare these to those obtained 
from the electron transport data in three different QD devices.
Two of the experimental samples (sample A and B) are lateral quantum dots
on a high-mobility 2DEG~\cite{ciorga00} while the third one (sample C) is
a vertical QD.\cite{oosterkamp} 
The samples A and B were manufactured on a high mobility 2DEG samples
with spin-polarized leads for electron transport measurements
in the spin blockade regime.
The data of the sample C was obtained in the Coulomb blockade.
The high mobility of samples chosen for comparison
is essential to reduce unpredictable effects of impurities and
disorder that make identification of signals of
physical phenomena difficult.

\begin{figure}[ht]
\includegraphics[width=0.99\columnwidth]{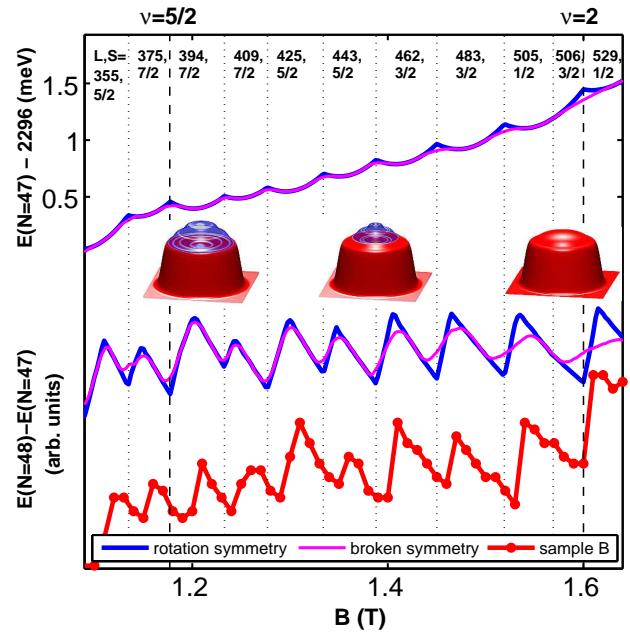}
\caption{(Color online)
Ground state energy of the $N=47$ quantum dot calculated
with the density-functional theory for the rotationally symmetric 
eigenstates of the angular momentum $L$ (blue curve) and for the
ground states in the symmetry-unrestricted approach (magenta curve). 
The corresponding chemical potentials $\mu(N=47\to 48)$ calculated from
the theory are shown in the lower panel together 
with experimental data from a lateral quantum-dot device (sample B).
Dashed lines correspond to the boundaries of the spin-droplet regime.
The insets show the fragmented spin and charge densities
of three of the corresponding states (cf. Fig. \ref{tippafigure}).
The strength of the parabolic confining 
potential of the quantum dot is $\omega_0=2\;{\rm meV}$ in the calculations.
}
\label{supplyfig1}
\end{figure}

We first address the problem of whether the electronic states
in the QD samples show any signs of broken rotational symmetry.
Inhomogeneities and impurities in QD devices may
break the rotation symmetry, and a Jahn-Teller type of mechanism
could be active if disorder alters significantly the shape of the
confining potential. As a result, the ground state transitions
with increasing magnetic field become continuous rather than discrete.
A signature of this type of symmetry breaking would
be a smoothing of the chemical potential.
Experimental data from a high-mobility lateral QD device
is of sufficiently good quality to test for the presence of 
symmetry breaking mechanisms. Figure~\ref{supplyfig1}
shows a comparison of the electron transport data to DFT calculations
with and without symmetry breaking. The data shows sharp increases
in the chemical potentials, which is consistent with discrete
transitions in the ground state.
Therefore, to a good approximation,
the rotational symmetry is preserved in high mobility samples,
and the angular momentum $L$ is a good quantum number.

\begin{figure*}
\includegraphics[width=0.8\linewidth]{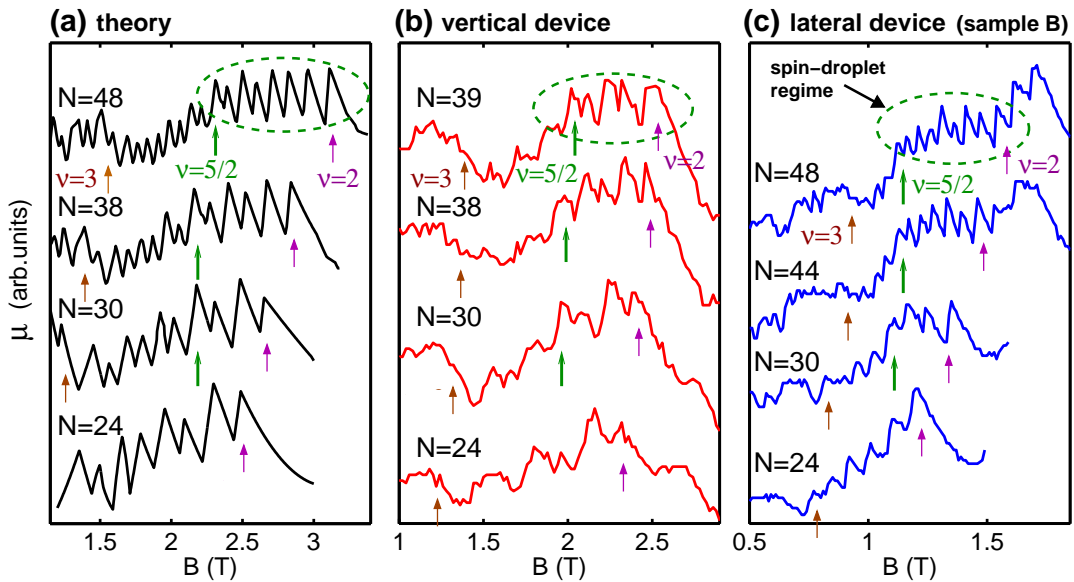}
\caption{(Color online)
Chemical potentials calculated with the density-functional theory 
(a) and measured
from a vertical (b) and lateral (c) quantum-dot devices
for various electron numbers. Both experiments show
the signal associated with the polarization of the second lowest Landau level
at $\nu=5/2$ in the peak position data
when $N\gtrsim 30$ in agreement with the theoretical result.
The confinement strength ranged from 2
to 4 meV depending on the electron number.
The data for the vertical device in (b) is courtesy of
L. Kouwenhoven,\cite{oosterkamp} and the
data for lateral device in (c) is
courtesy of A. S. Sachrajda.\cite{ciorga00}
}
\label{fig3}
\end{figure*}

The complete polarization of the SLL at $\nu={{5 \over 2}}$
is reflected in the energetics of the system.
The DFT calculations show that at $\nu={{5 \over 2}}$ there is a
step feature followed by a plateau region in the chemical potential.
Figure~\ref{fig3}
shows the DFT results for chemical potentials of
$N=24\ldots 48$ in comparison with the experiments.
The step in the chemical potential
is associated with the total polarization of the SLL in the DFT
calculations.
This feature can be found in all three samples we studied above $N\ge 30$,
as predicted by the theory.
Some models of QDs assume that the SLL is spin-polarized due to
the Zeeman effect.\cite{asta} This model does not, however, apply for 
the lateral and vertical QD devices examined in this work where the effect of
the Zeeman splitting is estimated to be only a few percent of
the spin splitting caused by the many-body interactions
(see Fig. \ref{supplyfig4}).

In the ${5 \over 2}>\nu> 2$ regime, the ground state energy is
approximately constant (see Appendix),
and the calculations show
a phase transition in the system where
two phases ($\nu=2$ and $\nu=3$)
coexist, and the size of the $\nu=2$ domain increases with the magnetic field.
The chemical potential does not continue to rise, but
instead, it is oscillating around a constant value until
$\nu=2$. This signature can be found in all the experimental samples
(see Fig. \ref{fig3}).
All electron transport data presented are thus consistent
with the theoretical picture that the ground states 
in the vicinity of $\nu={5\over 2}$ involve fragmented QH states.
We point out, however, that the results are sensitive
to the shape of the external potential, and the pairing
of the electrons may still occur if the potential is
sufficiently homogeneous, e.g., in large QDs, where
the second Landau level would acquire higher angular momentum.

\begin{figure*}
\includegraphics[width=0.7\linewidth]{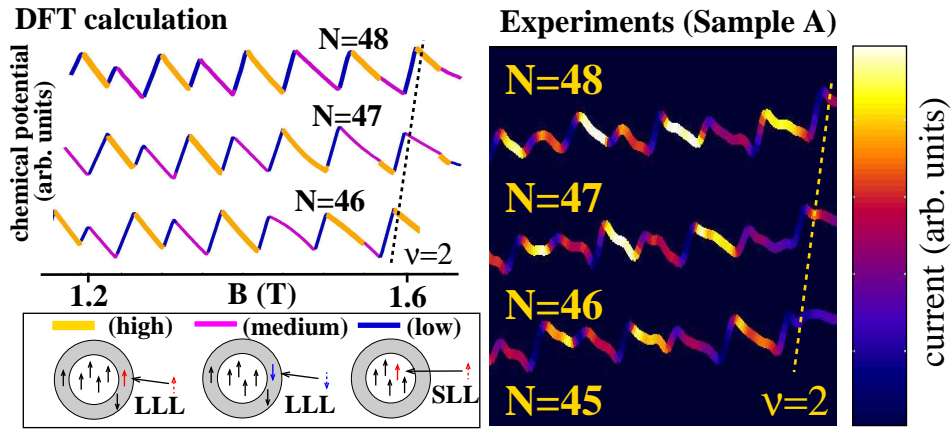}
\caption{(Color online) Checkerboard pattern of transport current
in density-functional theory (left panel) and spin-blockade
experiments (right panel). The lowest current densities correspond
to electron transport via states in the second-lowest
Landau level, near the core of the quantum dot.
The current density in experiments has been amplified in high
magnetic fields with a linear function to compensate
for the general attenuation of the signal.
}
\label{supplyfig5}
\end{figure*}

Spin polarization of the leads
is commonly used to create a current that
depends on the orientation of the electron spin, which passes
through the device. In the case of the two lateral QDs
in our analysis, the electrons enter the QD from spin-polarized
magnetic edge states of the 2DEG through tunneling barriers.
Coulomb blockade lifts when the energies of the many-body states corresponding
to $N$ and $N+1$ electrons are equal. The tunneling current depends
then on the coupling between the wave function in the QD
and the electronic states in the external leads.
The lowest Landau level orbitals are at the edges of the QD,
and the coupling is stronger to the leads compared to
the second lowest Landau level orbitals that are close to the center
of the QD. Due to polarization of the leads, their coupling 
to electron states with spin parallel to the external
polarization is higher than the coupling of
spins antiparallel to the external polarization.
This spin dependence in the transport has been shown to lead to
a characteristic checkerboard pattern
of current densities through
QDs.\cite{ciorga00,klitzing01,hitachi,kupidura,rogge}
Our DFT results are consistent with such
transport currents in the spin blockade regime (Fig.~\ref{supplyfig5}).
The polarization of the SLL in the ${{5 \over 2}} \ge \nu\ge 2$ regime
would be in contrast to the model presented in Ref.~\onlinecite{klitzing01}.
A consequence of this is that the transport current via SLL orbitals
should show no checkerboard pattern in this regime since the spins are
always parallel to the polarization of the leads. This could be
tested with high-accuracy spin blockade spectroscopy 
which would be able to detect small changes in the weak tunneling
currents through the SLL orbitals.

\section{Fragility of the $\nu={{5 \over 2}}$ quantum Hall state}
\label{sec:conclusions}

The $\nu={{5 \over 2}}$ state is one of the
most fragile QH states.
It is observed only in high-mobility 2DEG samples
as the paired electron state
may break down in the presence of impurities.
These induce a non-uniform potential that, in the light of
results in this work, may lead to its instability.
Our findings are thus
in line with those obtained by Chklovskii and Lee who predicted that in
the presence of long-range disorder in the 2DEG,
incompressible integer filling
factor regions form that are separated by domain walls.\cite{chklovskii}
These structures are analogous to the fragmented 
QH states that we find in QDs.
Structures reminiscent of domain walls have been observed with
scanning-probe imaging techniques in a perturbed
QH liquid.\cite{tessmer}

Analogous instability of QH states
may also occur in other geometries where the electrons are
not strictly confined in all directions, such as
in high-mobility 2DEG samples in the vicinity of constrictions.
One indication of this may be the observed fragility of the
$\nu={5 \over 2}$ state in narrow quantum point contacts.\cite{miller}
Proposed tests\cite{sternbondersonfeldman}
for the non-abelian properties of quasi-particle excitations
of  $\nu={{5 \over 2}}$ QH state make use of finite geometries
and multiple constrictions to generate interference among
tunneling paths. A possible fragmentation of the $\nu={5 \over 2}$
QH state close to the boundaries, which would lead to the
instability in such geometries, is still an open question
that requires further analysis of the
effects of the confinement.
While recent experiments on the quasi-particle tunneling,\cite{radu}
shot noise generated by partitioning edge currents,\cite{dolev}
and interferometric measurements of QH edge excitations
\cite{willett2} 
of the $\nu={5 \over 2}$ QH state
show results, which are all consistent with the unusual quasi-particle
charge $e^*={1 \over 4}$ of the paired electron state,
the particle statistics of the excitations remains to be confirmed.
Possible fragmentation of QH states
in narrow constrictions needed for quasi-particle interferometry
adds another challenge in this long quest to confirm the possible
non-abelian characteristics of the $\nu={5\over 2}$ state.

To conclude, we have shown theoretical evidence that
electron pairing is possible in small QH droplets
in quantum dots at $\nu={5 \over 2}$,
provided that the half-filled Landau level can
acquire sufficiently high angular momentum. However,
our calculations indicate that in parabolic external confining potentials
the paired electron state breaks down leading to fragmentation
of charge and spin densities.
We find indirect evidence of such fragmentation in several experiments
but point out that our results can be tested by direct measurements of
the spatial dependence of spin and charge densities in different
geometries and experimental setups.

\begin{acknowledgments}
We gratefully acknowledge valuable discussions with
A. S. Sachrajda, M. Ciorga, S. M. Reimann, and L. Kouwenhoven,
and thank Jaakko Nissinen for calculating the Pfaffian interaction matrix
elements. This work was supported by the EU's Sixth Framework Programme
through 
the Nanoquanta Network of Excellence (NMP4-CT-2004-500198),
the Academy of Finland, and the Finnish Academy of Science and Letters
through the Viljo, Yrj{\"o} and Kalle V{\"a}is{\"a}l{\"a}
Foundation.
\end{acknowledgments}

\section{Appendix: accuracy of numerical methods}

\begin{figure}
\includegraphics[width=0.99\columnwidth]{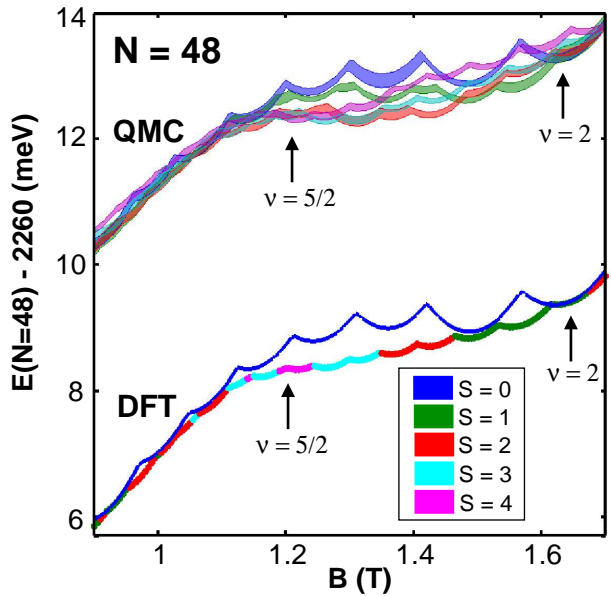}
\caption{(Color online)
Comparison of the ground state energies for given total spin $S$ in
the density-functional theory (DFT) 
and the quantum Monte Carlo (QMC) method. The number of electrons $N=48$.
The line widths in the QMC denote the statistical error in the results.
Only the ground state and the $S=0$ state are shown in the DFT result. 
The strength of the parabolic confining potential of the quantum dot 
is $\omega_0=2\;{\rm meV}$ in the calculations.
}
\label{supplyfig2}
\end{figure}

The electron correlations play an important role in the 
structure of fractional QH states.
To test for the accuracy of the DFT method in the spin-droplet
regime, we compare the energies of different spin polarization states
between the DFT and the QMC in the ${{5 \over 2}} \ge \nu \ge 2$ regime.
The results for a 48-electron QD are shown in Fig.~\ref{supplyfig2}.

Both methods show the spin-droplet structure with a comparable
energy benefit in the polarization $\delta\approx 0.5$ meV for $S_{\rm max}=4$.
The QMC method estimates that the maximum size of the spin droplet
is $N_{\rm SD}=7$ compared to $8$ in the DFT.
Given the typical statistical error of $\pm 0.05$ meV in the QMC 
results, the overall agreement between the methods is excellent.
This test indicates that the DFT method captures the essential 
many-body physics of the spin-droplet formation and
gives accurate results for the ground states. The DFT method was 
subsequently used in the calculation of the chemical potentials
of large QDs, which can be compared to the transport experiments
in the spin blockade regime.

The DFT method predicts some non-compact states outside the spin-droplet
regime, e.g., $L=375$, $S=7/2$ state as shown in Fig.~\ref{supplyfig1}.
This state has one spin-down electron in the SLL with $l=0$.
Emergence of non-compact states is a manifestation of the degeneracy of
the single-particle states near Fermi-level. However, they are rare in the DFT 
and occur only at magnetic fields below the polarization of the SLL.
Detailed analysis of these states with the QMC goes beyond the scope of
the present work and is left for future research.

\end{document}